\documentclass[preprint,showkeys, preprintnumbers,amssymb]{revtex4}

\usepackage{graphicx}
\graphicspath{{figs/}}
\begin{document}
\title{Thermal contraction in silicon nanowires at low temperatures}
\author{Jin-Wu~Jiang}
\thanks{Electronic mail: phyjj@nus.edu.sg}
	\affiliation{Department of Physics and Centre for Computational Science and Engineering,
 		     National University of Singapore, Singapore 117542, Republic of Singapore }
\author{Jian-Sheng~Wang}
	\affiliation{Department of Physics and Centre for Computational Science and Engineering,
     		     National University of Singapore, Singapore 117542, Republic of Singapore }
\author{Baowen~Li}
        \affiliation{Department of Physics and Centre for Computational Science and Engineering,
                     National University of Singapore, Singapore 117542, Republic of Singapore }
        \affiliation{NUS Graduate School for Integrative Sciences and Engineering,
                     Singapore 117456, Republic of Singapore}
\date{\today}
\begin{abstract}
The thermal expansion effect of silicon nanowires (SiNW) in [100], [110] and [111] directions with different sizes is theoretically investigated. At low temperatures, all SiNW studied exhibit thermal contraction effect due to the lowest energy of the bending vibration mode which has negative effect on the coefficient of thermal expansion (CTE). The CTE in [110] direction is distinctly larger than the other two growth directions because of the anisotropy of the bending mode in SiNW. Our study reveals that CTE decreases with an increase of the structure ratio $\gamma=length/diameter$, and is negative in whole temperature range with $\gamma=1.3$.
\end{abstract}

\keywords{thermal contraction, silicon nanowire, phonon mode}
\maketitle

\section{introduction}
Silicon nanowires (SiNW) can be well produced in different directions with different diameters.\cite{Morales, ZhangYF, DuanXF, Holmes, MaDDD, WuY} It is a promising candidate for nano-device, and has attracted a lot of interest.\cite{ChungSW, CuiY2001A, CuiY2001B, Gudiksen, Wong, Pan, HuangY2002, CuiY2003, XiangJ, TimkoBP, HuangY2001, TianB, HuangY2001B} The SiNW can serve as basic building blocks for electrically based sensor\cite{ChungSW, CuiY2001A, CuiY2001B, Gudiksen}, field effect transistor\cite{Wong, Pan, HuangY2002, CuiY2003, XiangJ, TimkoBP} and logic gates\cite{HuangY2001}, photovoltaic device\cite{TianB}, and functional networks\cite{HuangY2001B}. The possible stable configurations of the SiNW have been investigated with density-functional tight-binding simulations by Zhang {\it et~al.} in Refs.~\onlinecite{ZhangRQ}. They found that the stability of SiNW is determined by the competition between the minimization of the surface energy and the minimization of the surface-to-volume ratio. As one of the important thermal properties in SiNW, the thermal expansion plays an important role for these applications in the electronic or thermal devices. The thermal expansion effect can be studied in the classical molecular dynamics.\cite{ZhaoH}

In this paper, we first analyze various phonon vibrational modes in SiNW from the Tersoff potential\cite{Tersoff} implemented in the ``General Utility Lattice Program" (GULP).\cite{Gale} Some important features of phonon modes in SiNW of different directions and with different structure ratio $\gamma=length/diameter$ are discussed. We then investigate the coefficient of thermal expansion (CTE) of SiNW in [100], [110] and [111] growth directions by the nonequilibrium Green's function approach\cite{JiangJW}, which is a quantum mechanical method and automatically includes contribution of all phonon modes. We find that at low temperatures, all SiNW studied have thermal contraction effect and the CTE in SiNW [110] is the largest one among the three directions. The CTE decreases rapidly with increasing structure ratio $\gamma$, and is negative in whole temperature range with $\gamma=1.3$.

\section{results and discussion}
\begin{figure}[htpb]
  \begin{center}
    \scalebox{1.2}[1.2]{\includegraphics[width=7cm]{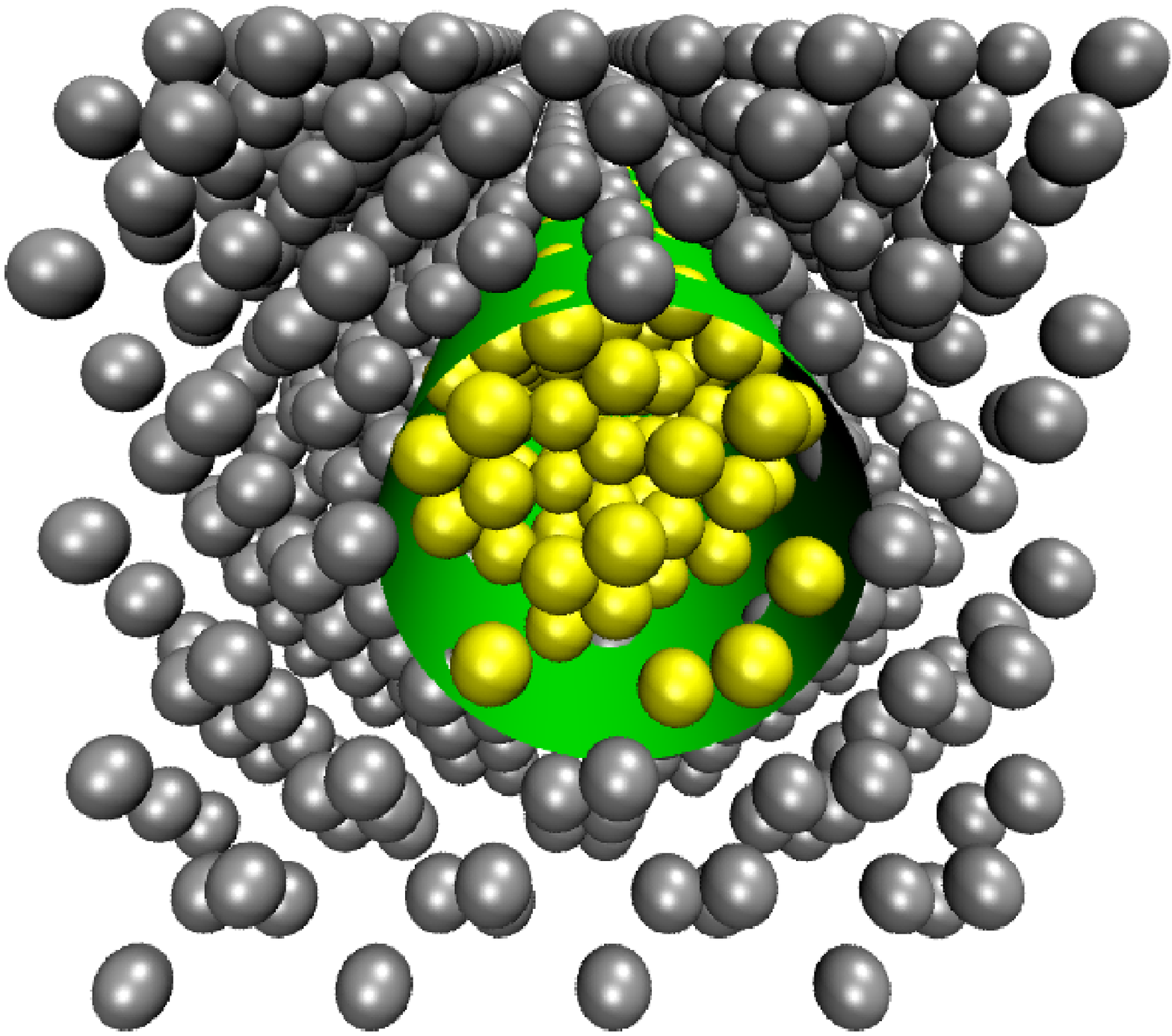}}
  \end{center}
  \caption{(Color online) Use a virtual cylinder to cut a SiNW from bulk Silicon.}
  \label{fig_cfg}
\end{figure}

Fig.~\ref{fig_cfg} demonstrates that a SiNW with length and diameter ($L$, $D$) can be cut from the bulk Si crystal by using a virtual cylinder with structure parameters ($L$, $D$).\cite{VoT} We can cut SiNW in different growth directions: [100], [110] and [111], by controlling the direction of the virtual cylinder. The SiNW displayed in the figure is in [100] growth direction with ($L$, $D$)=(3, 1) nm. As discussed in Refs.~\onlinecite{Markussen}, it is not important to include H-passivation on the surface for thermal property of SiNW. We adopt this approximation which may introduce some small deviation. To depress this small deviation, we do not use these atoms on the surface and boundaries. It seems that Si atoms in the figure do not distribute uniformly in the figure. That is because the structure is viewed from the upper front direction, which is better for illustrating a three-dimensional structure. Other configuration figures in this paper may also have this visual effect.

The frequencies and eigen vectors of phonon modes are obtained from the Tersoff interatomic potential for silicon implemented in the GULP. The Tersoff potential can well describe bonding in covalently bonded solids such as silicon and carbon system, and also includes the nonlinear interaction. Its accuracy has been confirmed by comparison with results from {\it ab initio} density functional theory.\cite{Markussen} In the investigation of CTE in the SiNW, we will use the clamped boundary condition, where one end (left, for example) is fixed and another end (right) is free. For consistency, we also apply this boundary condition in the study of phonon, which is different from Refs.~\onlinecite{Thonhauser2004, Peelaers}.

\begin{figure*}[htpb]
  \begin{center}
    \scalebox{1.0}[1.0]{\includegraphics[width=\textwidth]{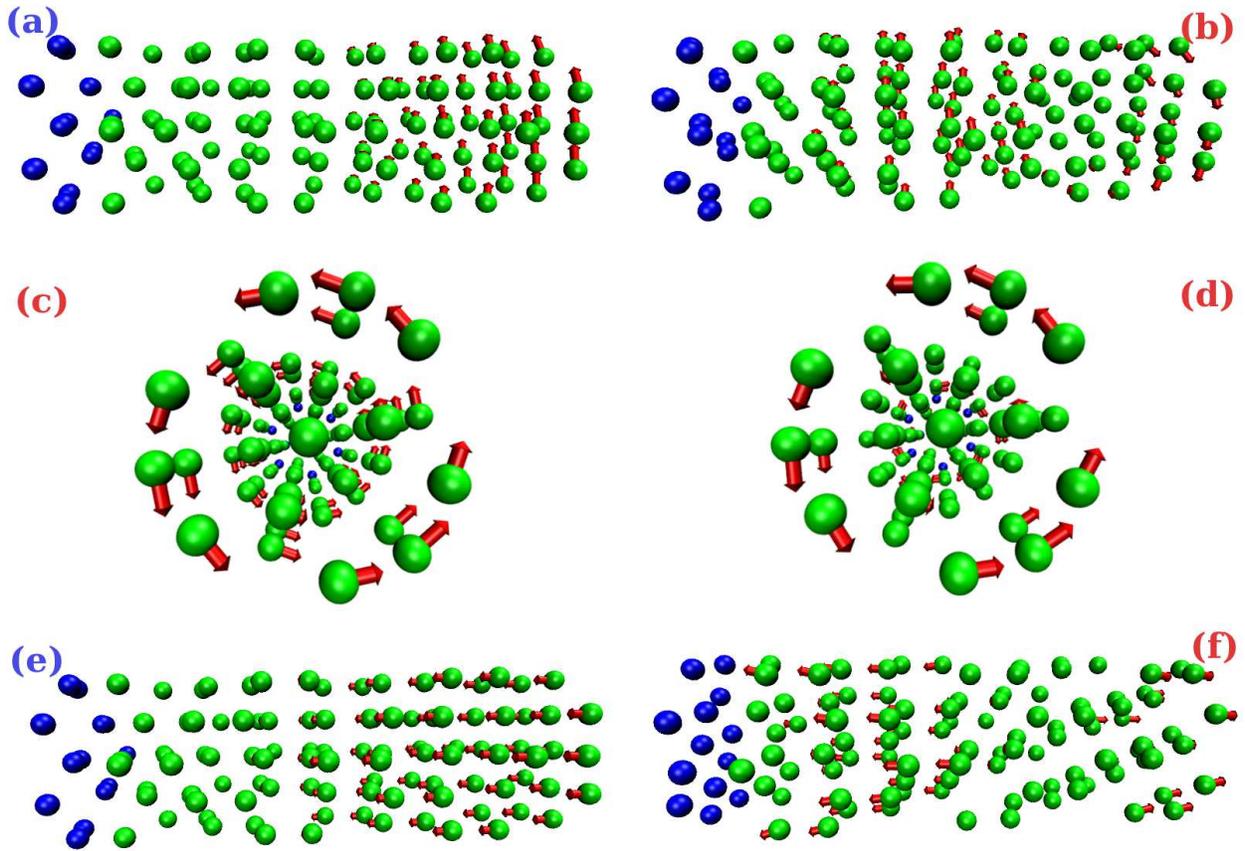}}
  \end{center}
  \caption{(Color online) The morphology of different phonon modes in SiNW. (a)/(b): the first/second order bending mode; (c)/(d): the first/second order twisting mode; (e)/(f): the first/second order longitudinal mode. }
  \label{fig_eigvec}
\end{figure*}

Fig.~\ref{fig_eigvec} shows the vibrational morphology of three important phonon modes for CTE in SiNW. In the figure the SiNW is in [111] growth direction with ($L$, $D$)=(3, 1) nm. The left-most 0.5 nm region (blue online) is fixed (clamped). Arrows (red online) attached onto each atom display the vibrational displacement for the atom.\cite{Thonhauser2009} The length of the arrow is not the exact value of vibrational displacement to show this displacement more clearly. It has been enlarged by a constant factor to all atoms. Panels (a) and (b) are the first and second order bending modes, which is a lateral mode with all atoms vibrate in the direction perpendicular to the axial direction. It is doubly degenerate due to two independent lateral vibrational directions. As the SiNW is clamped on the left, the vibrational displacement increases from left to right in both modes. In panel (a), the direction of the vibrational displacement does not change, while this direction changes in the second order mode in panel (b). The bending mode is actually a characteristic property for rod-like systems (eg. SiNW) or two-dimensional plane sheet (eg. graphene). Using its relation to the Young's modulus of the material, the bending mode has been used to detected the value of the Young's modulus of the low-dimensional systems such as single-walled carbon nanotubes\cite{Treacy} or graphene.\cite{JiangJW2009} For longer SiNW, this mode will gradually turn to the standard flexure mode in a rod-like systems. Panels (c) and (d) are the first and second order twisting modes. Because of its special rotating morphology, the vibrational displacement increases from inner to the surface. This is consistent with the vanishing surface stress condition in a rod.\cite{Thonhauser2004} In panel (c), the direction of the vibrational displacement is the same for all atoms, while the direction changes for the second order mode shown in panel (d). The twisting mode reflects the rod-like structure of the SiNW and is determined by the shear stiffness of the SiNW. It can serve to sense the tiny torsion movement of a nano device.\cite{Meyer} This mode can be used to extract the internal information of the SiNW, as it carries the information of the rigid rotational invariance symmetry.\cite{Mahan2004, Popov, JiangJW2006} Panels (e) and (f) are the first and second order longitudinal vibrational phonon modes. In very long SiNW, this mode is actually the well-known longitudinal acoustic mode.

\begin{figure}[htpb]
  \begin{center}
    \scalebox{1.2}[1.2]{\includegraphics[width=7cm]{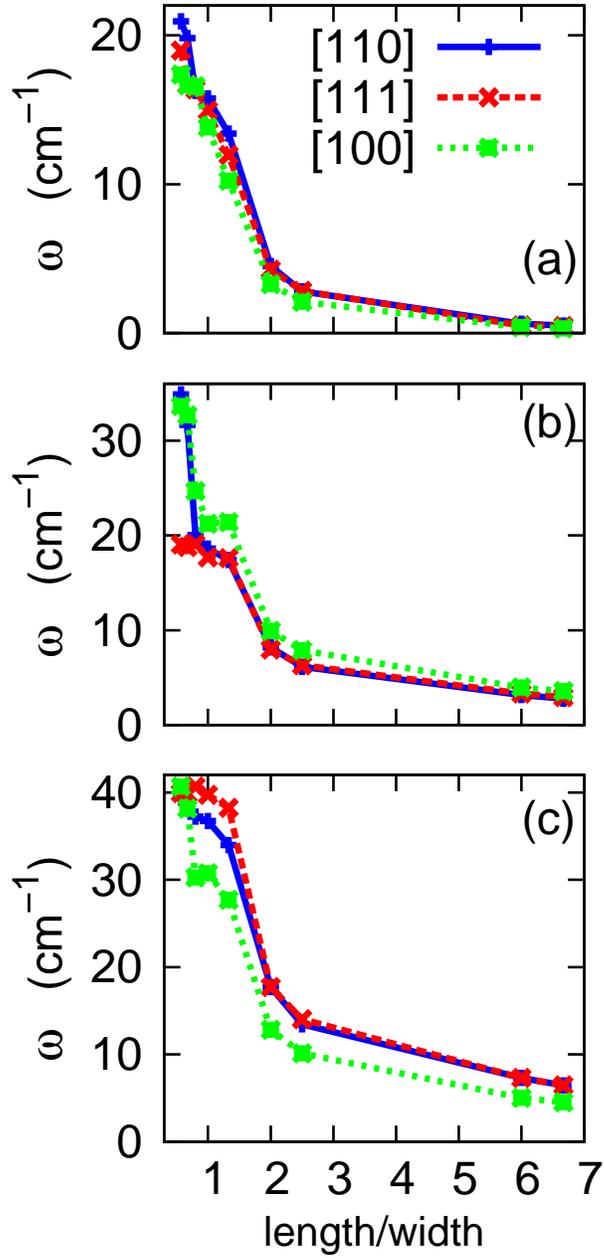}}
  \end{center}
  \caption{(Color online) The frequency of three lowest-energy phonon modes in SiNW with different size. (a). bending mode, (b). twisting mode, (c). longitudial mode.}
  \label{fig_mode}
\end{figure}

Fig.~\ref{fig_mode} shows frequencies of the three lowest-energy phonon modes of SiNW in different directions with different sizes. Frequencies for all phonon modes decrease with increasing structure ratio $\gamma=length/diameter$. This result indicates that longer SiNW are more flexible and elastic. As a result, SiNW with large $\gamma$ may be thermally more unstable, and possibly show a coiling-up behavior like a single-walled carbon nanotube.\cite{OuYang} Typically, we note that the frequency of bending mode in (a) is the lowest one compared with the other two modes shown in panels (b) and (c). This is because the bending mode turns to be the elastic flexure mode in the limit of $\gamma\rightarrow \infty$. An important feature of the flexure mode is its quadratic phonon spectrum: $\omega \propto k^{2}$. While the other acoustic phonon modes are linearly dependent on the wave vector. So the flexure mode (also bending mode) always has the lowest frequency. This figure also shows that the bending mode has lowest frequency in [100] direction and highest frequency in [110] direction with the same $\gamma$. So the SiNW in [100] direction are easiest to be bend, which may be due to its regular surface structure.

We calculate the CTE along the axial direction of the SiNW using the nonequilibrium Green's function method, which includes all phonon modes automatically and can be applied to very low temperatures as it is a quantum approach.\cite{JiangJW} In this method, firstly, the average vibrational displacement of atom $j$, $\langle u_{j}\rangle$, is calculated through: $\langle u_{j}\rangle=i\hbar G_{j}$, where $G_{j}$ is the one-point Green's function and can be calculated from its Feynman diagram expansion in terms of the nonlinear interaction. We consider the nonlinear interaction as,
\begin{eqnarray}
H_{n} & = & \sum_{lmn}\frac{k_{lmn}}{3}u_{l}u_{m}u_{n},
\end{eqnarray}
where the nonlinear force constant is extracted from the Tersoff potential in GULP by finite difference method. Then the CTE can be calculated through its definition,
\begin{eqnarray}
\alpha_{j}=\frac{d\langle u_{j} \rangle}{dT}\times\frac{1}{x_{j}},
\end{eqnarray}
where $x_{j}$ is the lattice position of atom $j$ along the axial direction. We can get the final value of CTE for SiNW by averaging $\alpha_{j}$ over all atoms.

\begin{figure}[htpb]
  \begin{center}
    \scalebox{1.4}[1.4]{\includegraphics[width=7cm]{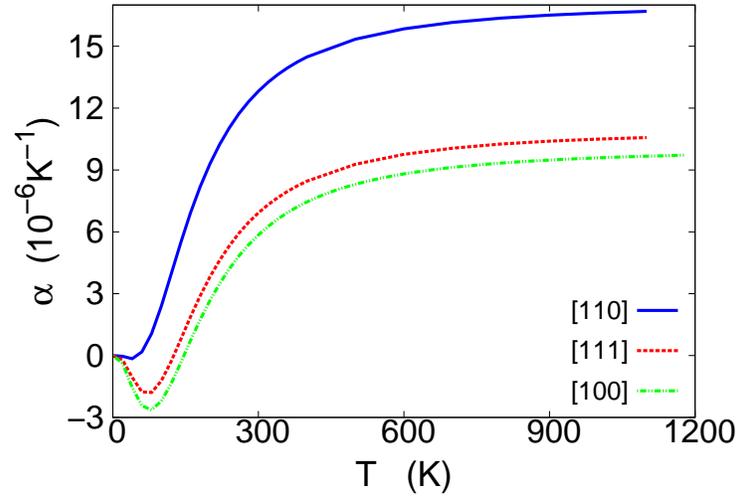}}
  \end{center}
  \caption{(Color online) CTE of SiNW in three directions [100], [110], and [111].}
  \label{fig_cte_direction}
\end{figure}

In Fig.~\ref{fig_cte_direction}, we compare the CTE in three different directions. All three SiNW have the same structure parameters ($L$, $D$) = (1, 4) nm. It shows that the CTE in all directions are negative in very low temperature region. Then it increases with further increasing temperature and reach a saturate value in high temperature limit. The saturate value is positive for these SiNW shown in the figure. We mention that before reaching the saturate value, all three curves still show obvious increasing behavior at about $T=500$ K, which implies that the Debye temperature in SiNW is roughly about 500 K. From this figure, we also learn that the CTE in [110] direction is distinctly larger than the other two directions. While the CTE in [100] direction is close yet a little smaller than that in [111].

The above behaviors for the CTE at different temperatures and in different directions can be understood from the various phonon modes in SiNW discussed in the previous part. Because of different vibrational morphology, these phonon modes have very different contribution to the CTE. The bending mode has negative effect on the CTE of the rod-like SiNW.\cite{Krishnan, Schelling} The twisting mode only takes up some degrees of freedom, without making any contribution to CTE. The longitudinal vibration mode has positive effect on the CTE. So the CTE is mainly determined by the competition between the bending mode and longitudinal mode. Since the bending mode has the lowest energy among all phonon modes, it is the only mode excited at low temperature, leading to negative CTE. With the increase of temperature $T$, the longitudinal mode will be excited gradually and competes with bending mode. This competition slows down the decreasing of CTE, and results in a minimum point. As the longitudinal mode becomes more and more important, CTE will increase with further increasing temperature. In high temperature limit, CTE reaches a saturate value after sufficient competition between all phonon modes. The positive saturate value in these SiNW implies the leading contribution of longitudinal mode for these SiNW. Also the bending mode is anisotropic with the frequencies in [110], [111] and [100] directions as 53.1, 41.2, 39.3 cm$^{-1}$, respectively. This anisotropic property leads to the anisotropic effect of CTE, i.e. $\alpha_{[110]}>\alpha_{[111]}>\alpha_{[100]}$.

\begin{figure}[htpb]
  \begin{center}
    \scalebox{1.4}[1.4]{\includegraphics[width=7cm]{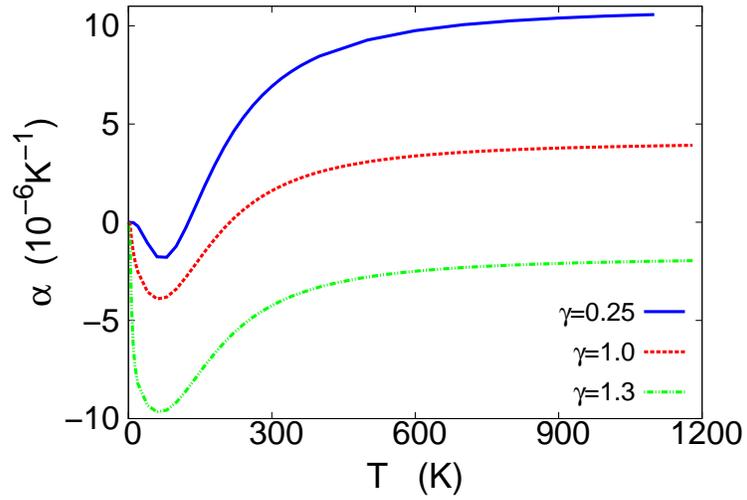}}
  \end{center}
  \caption{(Color online) CTE depends on $T$, in [111] SiNW with different structure ratio $\gamma=length/diameter$.}
  \label{fig_size111}
\end{figure}

We now turn to the discussion of the size dependence for CTE in SiNW [111]. There are similar effects for the other two directions. Fig.~\ref{fig_size111} shows the dependence of CTE on the structure ratio $\gamma$ in whole temperature range. The value of CTE decreases quickly with increasing $\gamma$. This is because the frequency of bending mode decreases quickly with increasing $\gamma$ (see Fig.~\ref{fig_mode}). As a result, the negative effect from bending mode on CTE increases. For SiNW with $\gamma=0.25$, the transition of CTE from negative to positive happens around 120 K, and CTE reaches a saturate value of $10\times 10^{-6}$ K$^{-1}$ in high temperature limit. A large positive saturate value of CTE indicates that the longitudinal mode makes dominant contribution rather than the bending mode in this SiNW. The transition temperature increases from 120 K to 270 K with the change of $\gamma$ from 0.25 to 1.0. The saturate value of CTE for SiNW with $\gamma=1.0$ decreases for more than 50$\%$ compared with that of $\gamma=0.25$. This result is reasonable and has also been observed in single-walled carbon nanotubes,\cite{JiangJW} which has similar one-dimensional rod-like structure. It seems to be a common phenomenon in one-dimensional system that the CTE is much smaller in longer and thinner structures. When $\gamma=1.3$, CTE is negative in whole temperature range, which indicates that the SiNW with these structure ratio undergoes thermal contraction at all temperatures. For these type of SiNW, the bending mode has played an dominant role over the longitudinal mode.

\section{conclusion}
To summarize, we have studied the thermal expansion of SiNW in different growth directions by the nonequilibrium Green's function method. We find that at low temperatures, all SiNW studied exhibit thermal contraction effect due to the lowest energy of the bending mode which has negative effect on CTE. The CTE in SiNW [110] is larger than the other two directions, because of higher frequency of bending mode in SiNW [110]. The CTE decreases quickly with increasing of the structure ratio $\gamma$, and is negative at all temperatures with $\gamma=1.3$.

\section{Acknowledgements}
The work is supported in part by a Faculty Research Grant of R-144-000-257-112 of NUS, and Grant R-144-000-203-112 from Ministry of Education of Republic of Singapore, and Grant R-144-000-222-646 from NUS.

\pagebreak

\end{document}